\documentclass[10pt,conference]{IEEEtran}

\usepackage{amsmath,amssymb,amsfonts}
\usepackage{xcolor}
\usepackage{graphicx}
\usepackage{float}
\usepackage{stfloats}
\usepackage[noend,ruled]{algorithm2e}
\usepackage[singlelinecheck=false,justification=justified,font=footnotesize,labelsep=period]{caption}

\begin{document}

\title{Adaptive Learning for\\Quantum Linear Regression}

\author{Costantino Carugno\inst{1,2}
\and Maurizio Ferrari Dacrema\inst{1} 
\and Paolo Cremonesi\inst{1}
}

\author{\IEEEauthorblockN{Costantino Carugno}
\IEEEauthorblockA{Politecnico di Milano}
\and
\IEEEauthorblockN{Maurizio Ferrari Dacrema}
\IEEEauthorblockA{Twentieth Century Fox\\
Springfield, USA\\
Email: homer@thesimpsons.com}
\and
\IEEEauthorblockN{James Kirk\\ and Montgomery Scott}
\IEEEauthorblockA{Starfleet Academy\\
San Francisco, California 96678--2391\\
Telephone: (800) 555--1212\\
Fax: (888) 555--1212}}

\author{
    \IEEEauthorblockN{
        Costantino Carugno\IEEEauthorrefmark{1}\IEEEauthorrefmark{2}, Maurizio Ferrari Dacrema\IEEEauthorrefmark{2}, Paolo Cremonesi\IEEEauthorrefmark{2}
    }
    \IEEEauthorblockA{\IEEEauthorrefmark{1} \textit{VTT Technical Research Center, Finland}}
    \IEEEauthorblockA{\IEEEauthorrefmark{2} 
    \textit{Politecnico di Milano, Italy}}
}

\maketitle
\IEEEpeerreviewmaketitle

\begin{abstract}
The recent availability of quantum annealers as cloud--based services has enabled new ways to handle machine learning problems, and several relevant algorithms have been adapted to run on these devices. In a recent work, linear regression was formulated as a quadratic binary optimization problem that can be solved via quantum annealing. Although this approach promises a computational time advantage for large datasets, the quality of the solution is limited by the necessary use of a precision vector, used to approximate the real--numbered regression coefficients in the quantum formulation. In this work, we focus on the practical challenge of improving the precision vector encoding: instead of setting an array of generic values equal for all coefficients, we allow each one to be expressed by its specific precision, which is tuned with a simple adaptive algorithm. This approach is evaluated on synthetic datasets of increasing size, and linear regression is solved using the D--Wave Advantage quantum annealer, as well as classical solvers. To the best of our knowledge, this is the largest dataset ever evaluated for linear regression on a quantum annealer. The results show that our formulation is able to deliver improved solution quality in all instances, and could better exploit the potential of current quantum devices.
\end{abstract} 

\begin{IEEEkeywords}
Adaptive Learning, Linear Regression, Quantum Annealing, Quantum Machine Learning
\end{IEEEkeywords}

\section{Introduction}
\label{sec:intro}
Linear regression \cite{book} is a simple and common machine learning algorithm used to perform predictive analysis \cite{ML}. Recently, it was shown that solving a regression problem on a quantum annealer, could provide a computational speed-up up to $2.8x$ for large datasets \cite{date}. However, this case study was aimed towards evaluating the scaling of the formulation, and the synthetic dataset was generated in a way that the quantum annealer could find exactly the specific real--valued regression coefficients. Since in practical scenarios the regression coefficients are unknown, a proper encoding of real--valued numbers into binary strings is a crucial challenge that quantum annealers need to face, in order to provide a high quality solution. Currently, due to hardware limitations, the limited precision available on current quantum devices prevents a refined reconstruction of the coefficients, thus constraining the evaluation of datasets with many features. In this work we propose a precision vector formulation suited for adaptive learning via iterative sampling (Sec. \ref{sec:algo}), and we test this algorithm on datasets of increasing size (Sec. \ref{sec:eval}). Our analysis shows that our adaptive algorithm is able to better approximate the coefficient values, compared to a fixed precision baseline, thus yielding improved solutions to the regression problem.

\section{Linear Regression}
\subsection{Problem Formulation}
The typical Linear Regression considers a dataset $X$, made of $N$ entries (rows) and $d$ features (columns), with an added bias feature in order to allow the model to capture the constant term, so that $X \in \mathbb{R}^{N \times D}$, with $D = d+1$. Each entry of the dataset is coupled to a real--valued label which serves as independent variables for the regression tasks, forming a vector $Y \in \mathbb{R}^{N}$.
The task of a linear regression model is to find a linear relationship of the features that predicts the target labels from the entries:
\begin{equation}
    y = w_0 + w_1 x_1 + w_2 x_2 + \dots + w_d x_d
\end{equation}
The regression coefficients represent the weights of the features, and are arranged in a solution vector $w \in \mathbb{R}^D$, so that the problem can be stated in matrix form as:
\begin{equation}
    Y = Xw
\end{equation}

\subsection{Classical Solution}
\label{sec:class}
The classical procedure of solving a linear regression problem is known as Ordinary Least Squares (OLS), and it relies in minimizing the squared Euclidean distance between the predicted values and the target labels:
\begin{equation} 
\min_{w \in \mathbb{R}^D} E(w) = || Xw - Y ||^2
\label{eq:ols}
\end{equation}
It is well known that, cast in such a way, this is one of the few machine learning problems that has a closed--form solution. To find the minimum, differentiate $E(w)$ with respect to $w$ and set the gradient equal to zero:
\begin{equation}
\nabla E(w) = 2X^T(Xw-Y) = 0 
\end{equation}
From which the solution coefficient vector follows:
\begin{equation}
    w = (X^T X)^{-1} X^T Y
\end{equation}
It is worth noting that there are scenarios when the computation using the direct formula is unfavourable, i.e. in large datasets, the computational cost of calculating $(X^T X)^{-1}$ becomes a significant bottleneck, as matrix inversion is typically an $O(d^3)$ operation, where $d$ is the number of features. In this case, one might make use of approximate methods, such as gradient descent, to adjust the weights iteratively by applying an update rule:
\begin{equation}
    w \rightarrow w - \alpha \nabla E(w),
\end{equation}
where $\alpha$ is the learning rate, a hyperparameter that determines the size of the step taken in each iteration.

\subsection{Quantum Solution}
\label{sec:quantum}
In order to be solved by quantum annealing devices, the linear regression problem needs to be stated in the form of a Quadratic Unconstrained Binary Optimization (QUBO) problem \cite{lewis}. In general, QUBO problems encode linear and quadratic relationship among variables in a matrix, so that the optimization problem can be represented as:
\begin{equation}
    \min_{z \in \mathbb{B}^M} z^T A z + z^T b,
    \label{eq:qubo}
\end{equation}
where
$z \in \mathbb{B}^M$ is a binary vector,
$A \in \mathbb{R}^{M \times M}$ is the QUBO matrix, and
$b \in \mathbb{R}^M$ is the QUBO vector -- i.e. the diagonal terms of the QUBO matrix.

The linear regression OLS formulation in Eq. \ref{eq:ols} is obviously a quadratic equation, and it can be expanded as:
\begin{equation}
    \min_{w \in \mathbb{R}^D} E(w) = || Xw - Y ||^2 = w^T X^T X w - 2 w^T X^T Y + Y^T Y
    \label{eq:quad}
\end{equation}

However, this equation does not directly fit the QUBO form because it involves real--valued weights, $w$, whereas QUBO requires binary variables. To adapt this to a QUBO--compatible form, each real--valued weight, $w_i$, is approximated using a linear combination of binary variables, and introducing a fixed precision vector, $p = (p_1, p_2, \ldots, p_K)^T$. In particular, each weight, $w_i$, is represented by $K$ binary variables $\hat{w}_{ik}$ which act as selectors of the entries of $p$, such that:
\begin{equation}
    w_i = \sum_{k=1}^K p_k \hat{w}_{ik}
    \label{eq:weight}
\end{equation}
where $p_k$ denotes the $k^{th}$ entry of $p$. To organize this into a matrix form suitable for QUBO, the binary weights are arranged into a vector $\hat{w} = (\hat{w}_{11}, \ldots, \hat{w}_{1K}, \ldots, \hat{w}_{D1}, \ldots, \hat{w}_{DK})^T$. A precision matrix, $P$, is then constructed using the Kronecker product of the identity matrix, $\mathbb{I}_{D}$, with the transpose of the precision vector, $p^T$. The weights $w \in \mathbb{R}^{D}$ can thus be expressed as:
\begin{equation}
w = {P} \hat{w}, \quad {P} = \mathbb{I}_{D} \otimes p^T
\label{eq:w}
\end{equation}
where $P \in \mathbb{R}^{D \times D\cdot K}$. This reformulation allows the linear regression problem to be handled, by quantum annealing devices by converting it into a binary optimization framework, making it compatible with the computational capabilities of these devices. In fact, by substituting the weights defined in Eq. \ref{eq:w}, a QUBO formulation for Eq. \ref{eq:quad} follows:
\begin{equation}
        \min_{\hat{w} \in \mathbb{B}^{D\cdot K}} E(\hat{w}) = \hat{w}^T P^T X^T X P\hat{w} - 2 \hat{w}^T P^T X^T Y,
\label{eq:quboreg}
\end{equation}
where the $Y^T Y$ term was discarded since constants do not affect optimization. Indeed, Eq. \ref{eq:quboreg} is equivalent to Eq. \ref{eq:qubo}, with:
\begin{equation*}
    z = \hat{w} \quad A = P^T X^T X P,\quad B = - 2 P^T X^T Y
\end{equation*}
\subsection{Adaptive Learning for the QUBO Formulation}
\label{sec:algo}
In this section, we describe a refined method for configuring the precision matrix, $P$, used in the QUBO formulation of linear regression, as previously introduced in Equation \ref{eq:w}. The original approach employs a single, static precision vector for all regression coefficients. This method, however, can lead to suboptimal representation of coefficients, due to the non--uniform significance and scale of different features in a dataset. To address this matter, we propose an adaptive strategy where each coefficient, $w_i$, is approximated by its own unique tailored precision vector, $\pi_i \in \mathbb{R}^K$. Each vector $\pi_i$ is composed of values specifically adjusted to optimally represent its corresponding coefficient, so that Eq. \ref{eq:weight} can be restated as:
\begin{equation}
    w_i = \sum_{k=1}^K \pi_{ik} \hat{w}_{ik},
\end{equation}
where $\pi_{ik}$ is $k^{th}$ entry of the precision vector relative to weight $w_i$. Each coefficient's precision vector, $\pi_i$, is dynamically adjusted through an iterative learning process, allowing for a more accurate representation. The new precision matrix, $P$, is thus constructed by combining the individual precision vectors using a set of binary matrices $\{\mathcal{I}^i\} \in \mathbb{B}^{D\times D}$. Each matrix $\mathcal{I}^i$ has all entries equal to zero except for the $i^{th}$ entry on the diagonal index $i$, which is equal to one: $\{\mathcal{I}^i\} = \{\ \mathcal{I}^i_{ii} = 1\ \wedge\ \mathcal{I}^i_{jk} = 0,\ \forall i\forall j\forall k \in (0, \ldots, D) \}$.
This method ensures that each coefficient's representation is isolated and adjusted independently. The precision matrix in Eq. \ref{eq:w} is then expressed as a summation of Kronecker products:
\begin{equation}
   w = {P} \hat{w}, \quad P = \sum_i \mathcal{I}^i \otimes \pi_i,
\end{equation}
where $\pi_i$ is the precision vector of the $w_i$ regression coefficient. The QUBO formulation of linear regression in Eq. \ref{eq:quboreg} can make use of this precision matrix without any further modification.

To refine each $\pi_i$, we employ an iterative optimization algorithm, detailed in Algorithm \ref{algo:ada}. Initially, all entries in each precision vector are set using a predefined range and granularity, defined by a step size known as the  (\textit{rate}). During each iteration, the linear regression model is recalibrated using the current precision vectors, and the fit's quality is evaluated. The quality of the fit is calculated using the coefficient of determination $R^2$, which compares the residual sum of squares (RSS) against the total sum of squares (TSS) \cite{libro}:
\begin{equation}
    R^2 = 1-\frac{RSS}{TSS},
\end{equation} 
where $y_n \in Y$ are the target labels, $\Bar{y} = \frac{1}{n}\sum_n y_n$ is the target mean, $f_n$ are the predictions of the target, $RSS=\sum_n(y_n - f_n)^2$, and $TSS=\sum_n(y_n - \Bar{y})^2$. 
If the quality improves over the previous iteration, each $\pi_i$ is adjusted to more closely approximate the corresponding $w_i$, it is re--centered towards the weight sampled, and reduced in range and in granularity by decreasing the step--size of a factor \textit{rate\_desc}. Conversely, if the quality of the fit decreases, each $\pi_i$ needs to be allowed to explore a larger range, so it is enlarged in range and granularity, by increasing the step--size of a factor \textit{rate\_asc}, without re--centering. The sampling and rate scaling is repeated for a predefined number of iterations or until improvements plateau, as indicated by changes in $R^2$.

\begin{algorithm}[ht!]
\scriptsize 
\SetKwInOut{Input}{Input}
\SetKwInOut{Output}{Output}
\Input{X, Y, w\_init, rate, rate\_desc, rate\_asc, n\_iter}
\Output{w\_best, $R2$\_best}
\BlankLine
$w, w$\_$best \gets w$\_$init$\; 
$R2$\_$old, R2$\_$best \gets calculateR2(X,Y,w)$\;
\While{$n$\_$iter > 0$}{
    $P \gets getPrecisionMatrix(X,Y,w,rate)$\;
    $Q \gets getQUBO(X,Y,P)$\; 
    $w$\_$new  \gets solveQUBO(Q)$\;
    $R2$\_$new  \gets calculateR2(X,Y,w$\_$new)$\;
    \If {$R2$\_$new > R2$\_$old$}{
        $w \gets (w+w$\_$new)/2$\;
        $rate \gets rate / rate$\_$desc$\;
    }
    \Else{
        $rate \gets rate * rate$\_$asc$\;
    }
    \If {$R2$\_$new > R2$\_$best$}{
        $R2$\_$best \gets R2$\_$new$\;
        $w$\_$best \gets w$\_$new$\;
    }
$R2$\_$old \gets R2$\_$new$\;
$n$\_$iter \gets n$\_$iter - 1$\;
}
\caption{Adaptive algorithm for precision tuning}
\label{algo:ada}
\end{algorithm}

\section{Evaluation}
\label{sec:eval}
In order to rigorously evaluate the algorithm proposed in Sec. \ref{sec:algo}, we employ both classical and quantum computational solvers to process synthetic datasets, $X$, each consisting of $10^6$ datapoints, with number of features $D$ that range from $10$ to $88$. The feature values are generated using numpy's \textit{randn} function, which draws samples from a standard normal distribution. Correspondingly, the target set, $Y$, is generated such that the regression weights, are real--value numbers uniformly distributed in the interval $[0, 1]$, with an added small Gaussian noise term. The choice of using synthetic datasets instead of real--world data is motivated by the concern that, in order for linear regression to be effective and yield reliable predictions, there are important statistical assumptions that the datasets need to satisfy \cite{applied}:
\begin{itemize}
    \item No auto--correlation -- the residuals should not be correlated with each other    
    \item Multivariate normality -- the residuals (the differences between observed and predicted values) should be normally distributed
    \item Homoscedasticity -- the variance of residual errors should be consistent across all levels of the independent variables
    \item Low multicollinearity -- independent variables should not be too highly correlated with each other
\end{itemize}
Such controlled conditions help in isolating the performance of the algorithm from external variability and ensure that the results are only reflective of the different solvers utilized. Moreover, the number of entries, $N$, was chosen such that the runtime of both classical and quantum solvers are comparable \cite{date}. To the best of our knowledge, this is the largest dataset ever evaluated for linear regression on a quantum annealer. 

The linear regression task is solved using a variety of techniques: the closed--form (CF) method in Sec. \ref{sec:class}, stochastic gradient descent (SGD), simulated annealing (SA) and quantum annealing (QA) solvers, where the latter two solvers utilize the QUBO formulation of Sec. \ref{sec:quantum}. In this formulation the number of variables needed in the QUBO encoding is $M = D\cdot K$, where $D$ is the number of features and $K$ is the length of the precision vector. Furthermore, for the QA solver there is an added increase in variable number, due to the topological constraints caused by the limited connectivity of the device (embedding procedure). In this analysis, we employ the D--Wave Advantage quantum annealer, which has $5,640$ qubits, but only $40,484$ couplers, which constrain its capacity to accommodate large models. We observe that, by limiting the precision vector to its possible shortest length of $2$ values, the largest dataset that can be embedded contains $88$ features, after which the embedding procedure fails. This is due to the fact that, since the QUBO matrix is dense, its evaluation requires an all--to--all connectivity. In fact, the number of qubits employed grows from a theoretical $M = 88\cdot2=176$, before embedding, to a practical $M = 2,939$, after the heuristical embedding procedure, in order to account for all the original $M^2 = 176^2=30,976$ interactions. A similar constraint also applies for datasets with a lower number of features, although the precision vector can contain progressively more values. However, these cases are less of interest from the computational perspective, as the closed--form method performs quickly and effectively.

The linear regression evaluation is established on three key criteria:
\begin{itemize}
    \item Model accuracy -- Assessed via the coefficient of determination, $R^2$, as defined in Sec. \ref{sec:algo}
    \item Computational efficiency -- Runtime, as measured by time--to--solution (TTS)
    \item Scalability -- Evaluated by incrementally increasing the number of features
\end{itemize}

The results of model accuracy are shown in Table \ref{tab:reg}, for datasets of increasing feature size. The classical methods are executed on a local machine equipped with an Intel i$9$ $10900$K processor and $32$GB of RAM. The CF and SGD methods use the default settings of \textit{sci--kit learn} and are meant to serve as baselines. All SA and QA methods are executed with a number of samples $num\_samples=1000$. For the fixed annealing methods the problem is solved using a precision vector, which divides the range in equal parts, of the maximum allowed length for the annealer and the dataset processed. Their adaptive counterparts, use this vector as starting point, $w$\_$init$, of Algorithm \ref{algo:ada}. The adaptive hyperparameters, $rate\_asc$ and $rate\_desc$, are post--selected after a grid search. We observe a decisive improvement in model accuracy in the adaptive case for all instances evaluated, both for SA and QA solvers. It is worth noting that, although SA mostly outperforms QA in quality, there are instances where the reverse is true, such as for $D=15$ and $D=30$, or are of similar quality, like $D=10$. Being optimization algorithms, both methods are prone to several heuristics (annealing time, temperature parameter, number of samples, etc.), so it is unclear whether to attribute this result to a (un)fortunate sampling or to a (worse) better exploration of the solution space by the (SA) QA solver, without undergoing the tedious task of isolating and evaluating the impact of each heuristic involved.  Certainly, the difference in quality between SA and QA, both adaptive and non--adaptive, shows the impact of the embedding process and the downside of having to map the problem to a specific topology. However, although the increase in feature number is generally detrimental to the quality of QA, as more noisy resources are employed, the adaptive version shows consistent robustness. Moreover, we acknowledge the limitation of the QUBO encoding on solution quality, as even the SA--Ada and QA--Ada methods are unable to reach the quality of the CF and SGD methods. This problem can only be alleviated with longer and more refined precision vectors, as suggested by the results in the smallest instances, $D=10,15$. While, in the case of SA, this would only translate to a larger memory requirement and more values to be adapted in the precision, in the case of QA, this would be enabled only by a substantial larger number of connections among available qubits, which is a major engineering challenge that needs to be tackled.

\begin{figure}
    \centering
    \includegraphics[width=0.49\textwidth]{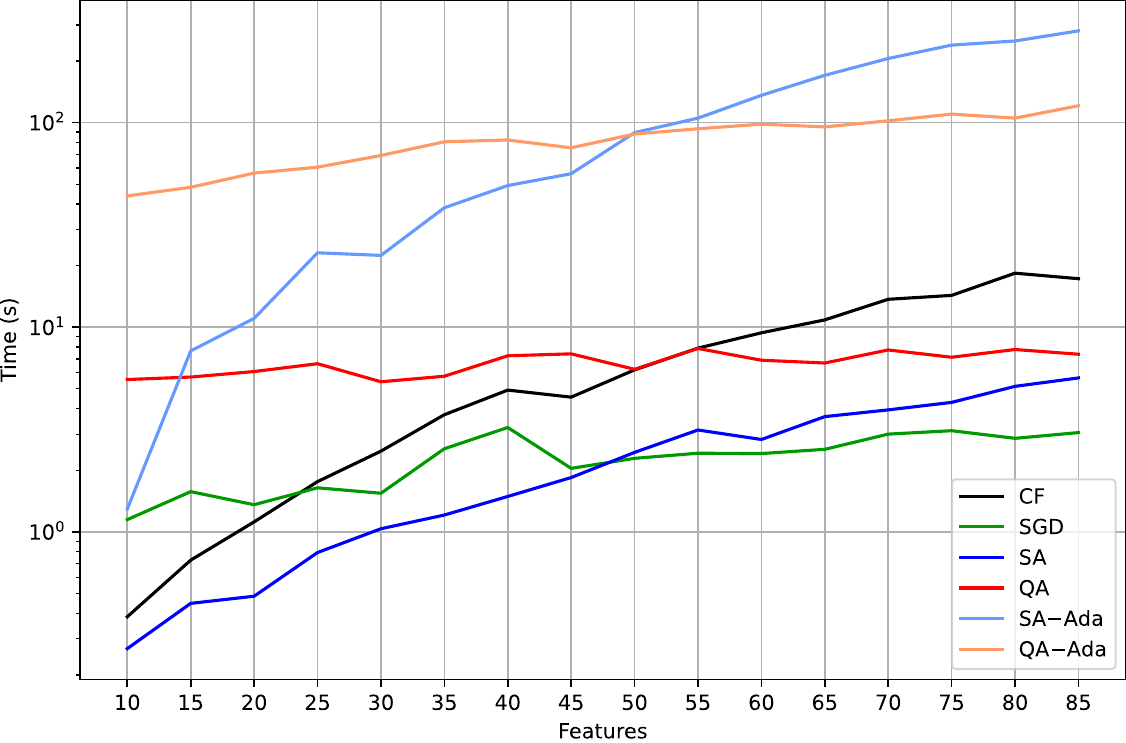}
    \caption{Logarithmic plot of time--to--solution (TTS) of linear regression evaluated on the datasets in Table \ref{tab:reg}.}
    \label{fig:time}
\end{figure}

In Fig. \ref{fig:time} the results for computational efficiency are presented, as time--to--solution (TTS) plotted on a logarithmic scale, for the datasets in Table \ref{tab:reg}. We observe that, as expected, CF exhibits a polynomial scaling, dominated by the matrix inversion operation, of complexity $O(d^3)$. Gradient descent is a better classical baseline for large datasets, being considerably faster, although its cost depends on the convergence of the method. Ordinary (batch) gradient descent requires the whole dataset to be evaluated and has cost per iteration $O(n\cdot d)$. Even faster is stochastic gradient descent (SGD), which
selects at random only one or few (mini--batches) datapoints to compute  the gradient, and has a cost per iteration $O(d)$. In our experiment, SGD performs exceptionally well, both on time and quality, which is to be expected, considering the assumptions on datasets generation. In fact, although it requires very few iterations before convergence, we do observe a moderate  increase in TTS as feature number increases. Regarding the QUBO methods, of which SA serves as baseline, it is important to underline that, since the solution vector $\hat{w} \in \mathbb{B}^{D\cdot K}$, the dimension of all the possible solutions obtainable grows exponentially, as $2^{D\cdot K}$. It is therefore crucial to have an optimization algorithm that can properly navigate this very large solution landscape. Our results show that SA is always faster than CF, and faster than SGD for small--size datasets. However, this gain in time does not yield an equivalent gain in quality, and if the accuracy is improved -- i.e. by applying the adaptive algorithm -- it comes at the expense of increased computational time. Moreover, in our experiment, SA has a similar scaling to CF, and is therefore unsuitable for large datasets. Differently, ordinary QA does not scale considerably over increasing dataset size. This behavior is consistent with the observation that the annealing process is constant in time, and the time--to--solution is effectively the result of loading, pre-- and post--processing the data. The adaptive quantum version, QA--Ada, displays a moderate increase in time cost if the dataset size is increased, comparable to the SGD scaling, while, on the same instance size, it shows a considerable time gap with its non--adaptive version, on a similar level to the difference that SA--Ada has with SA. This is due to the additional iterations required to optimize the precision vector, that we keep fixed in order to have a consistent benchmarking result. In fact, there is a similar trade--off as with SA: better solution quality can be traded for more computational time. In the quantum case, however, QA--Ada exhibits a more convenient scaling for large dataset, as it is faster than SA--Ada for datasets with size $50$ and more. This result is to be attributed to a combination of the efficiency of the quantum annealing process in exploring the solution space, coupled with the capability of the adaptive algorithm in fine--tuning the solution. 

Future works may improve on the methods presented here, as the code to reproduce these results is publicly available on GitHub \cite{repo}. 
\begin{table}[t]

\centering
    \begin{tabular}{|c|c|c|c|c|c|c|}
    \hline
    Features & CF & SGD    & SA     & SA--Ada & QA     & QA--Ada \\ \hline \hline
10       & 0.9629      & 0.9629 & 0.8426 & 0.9508      & 0.8426 & 0.9508      \\ \hline
15       & 0.9865      & 0.9865 & 0.8288 & 0.9618      & 0.8371 & 0.9622      \\ \hline
20       & 0.9861      & 0.9861 & 0.8484 & 0.9455      & 0.8390 & 0.9213      \\ \hline
25       & 0.9914      & 0.9913 & 0.9559 & 0.9711      & 0.8767 & 0.9391      \\ \hline
30       & 0.9916      & 0.9916 & 0.7969 & 0.9468      & 0.7811 & 0.9591      \\ \hline
35       & 0.9925      & 0.9925 & 0.8959 & 0.9574      & 0.8548 & 0.9213      \\ \hline
40       & 0.9922      & 0.9922 & 0.8963 & 0.9501      & 0.7416 & 0.9401      \\ \hline
45       & 0.9931      & 0.9931 & 0.9463 & 0.9852      & 0.8376 & 0.9290      \\ \hline
50       & 0.9951      & 0.9951 & 0.9690 & 0.9932      & 0.8639 & 0.9344      \\ \hline
55       & 0.9942      & 0.9942 & 0.9048 & 0.9643      & 0.7924 & 0.9355      \\ \hline
60       & 0.9940      & 0.9939 & 0.8901 & 0.9432      & 0.7196 & 0.9399      \\ \hline
65       & 0.9953      & 0.9953 & 0.9075 & 0.9542      & 0.7973 & 0.9254      \\ \hline
70       & 0.9961      & 0.9960 & 0.9572 & 0.9820      & 0.8241 & 0.9220      \\ \hline
75       & 0.9965      & 0.9965 & 0.9311 & 0.9713      & 0.7977 & 0.9118      \\ \hline
80       & 0.9968      & 0.9968 & 0.8984 & 0.9535      & 0.7699 & 0.9421      \\ \hline
85       & 0.9961      & 0.9961 & 0.9564 & 0.9857      & 0.7599 & 0.9356      \\ \hline
\end{tabular}

 \caption{Coefficient of determination ($R^2$) obtained by solving linear regression on datasets with $10^6$ datapoints of increasing feature size, using different solvers: closed--form (CF), stochastic gradient descent (SGD), ordinary and adaptive simulated annealing (SA, SA--Ada) and quantum annealing (QA, QA--Ada).}
 \label{tab:reg}
\end{table}

\section{Conclusions}
Quantum linear regression, formulated as a QUBO problem, has previously empirically demonstrated to be a faster alternative to its classical counterpart. However, the solution quality of the optimization is limited by currently available quantum devices, which struggle to embed the necessary precision. In this work, we present a new way of exploiting the precision vector formulation in order to refine the calculation via a simple adaptive algorithm, which was validated on increasingly larger datasets, that were synthetically generated to satisfy the requirements for a linear regression model. Problem instances of increasing feature size were solved using a variety of techniques, that served as different baselines: the closed--form (CF), stochastic gradient descent (SGD), simulated annealing (SA) and quantum annealing (QA), alongside with our proposed adaptive improvements: SA--Ada and QA--Ada.

In our experiment, we employ the largest quantum annealer currently available, D--Wave Advantage, and we confirm the previous result, obtained previously on D--Wave 2000Q, of a potential speedup of quantum linear regression (QA) over the closed--form approach (CF) for large datasets. To the best of our knowledge, this is the largest quantum linear regression so far executed on a real quantum device. Moreover, we focus our work on improving the quality of the solution, and we analyze our proposed adaptive tuning for the precision vector using both SA--Ada and QA--Ada. We find that, in both cases, our adaptive algorithm provides a consistently better solution quality, for a trade--off in computational time. However, several hurdles still constrain the quantum annealer to achieve the same capabilities of the classical solvers, of which the most critical is the number of qubits and the low connectivity among them, that ultimately limit the capabilities of executing large models or reaching a high precision. 

Several improvements to this work are possible, both on the classical and on the quantum side. On the classical side, the employment of a more sophisticated adaptive algorithm could allow a better precision or a shorter time, for example using a different initialization method for the weights, a faster update rule to converge to the optimal precision, and the inclusion of early stopping techniques. On the quantum side, a custom embedding for dense QUBOs, and the employment of advanced techniques such as reverse annealing or chain break mitigation may allow larger models and improved results.

Finally, we hope that future works may benefit from the results of this analysis for further evaluation and application of quantum linear regression in real--world scenarios.

\section{Acknowledgements}
We acknowledge the financial support from ICSC - ``National Research Centre in High Performance Computing, Big Data and Quantum Computing", funded by European Union – NextGenerationEU.

\bibliography{bib/refs}

\end{document}